\documentclass[a4paper,11pt]{article}
\usepackage{aaskaiid}
\usepackage{orcidlink}
\usepackage{rotating} % for sidewaystable (Referee 1 request: table landscape)
\title{The SKA-VLBI Perspective on Radio-Quiet AGN}
\ShortTitle{AGN accretion and ejection}

\author[1, \dagger]{Francesca Panessa
\orcidlink{0000-0003-0543-3617}}
\ShortName{Panessa et al.} % shortened name list for header 
\author[2,3 \dagger]{Tao An \orcidlink{0000-0003-4341-0029}}
\author[4, \dagger]{James Petley \orcidlink{0000-0002-4496-0754}}
\author[5, \dagger]{Ailing Wang \orcidlink{0000-0002-7351-5801}}
\author[6]{Ranieri~Diego Baldi \orcidlink{0000-0002-1824-0411}}
\author[7]{Ehud Behar
\orcidlink{0000-0001-9735-4873}}
\author[8,9]{Emmanuel Bempong-Manful \orcidlink{0000-0002-1727-1224}}
\author[1]{Gabriele Bruni \orcidlink{0000-0002-5182-6289}}
\author[10]{Ning Chang \orcidlink{0000-0002-8684-7303}}
\author[7]{Sina Chen \orcidlink{0000-0003-1586-3653}}
\author[10]{Lang Cui \orcidlink{0000-0003-0721-5509}}
\author[6]{Filippo D'Ammando \orcidlink{0000-0001-7618-7527}}
\author[6]{Marcello Giroletti\orcidlink{0000-0002-8657-8852}}
\author[11]{Magdalena Kunert-Bajraszewska\orcidlink{0000-0002-6741-9856}}
\author[12,13,14]{Sibasish Laha\orcidlink{0000-0003-2714-0487}}
\author[7]{Ari Laor\orcidlink{0000-0002-1615-179X}}
\author[15]{Ian McHardy\orcidlink{0000-0002-0151-2732}}
\author[16]{Eileen Meyer\orcidlink{0000-0002-7676-9962}}
\author[6]{Monica Orienti\orcidlink{0000-0003-4470-7094}}
\author[17,18]{Miguel P\'erez-Torres\orcidlink{0000-0001-5654-0266}}
\author[6]{Isabella Prandoni \orcidlink{0000-0001-9680-7092}}
\author[19,20]{Claudio Ricci\orcidlink{0000-0001-5231-2645}}
\author[21]{David R. A. Williams-Baldwin\orcidlink{0000-0001-7361-0246}}
\author[22]{Zsolt Paragi\orcidlink{0000-0002-5195-335X}}

\affiliation[1]{INAF - Istituto di Astrofisica e Planetologia Spaziali, via del Fosso del Cavaliere 100, Roma, I-00133, Italy}
\emailAdd{francesca.panessa@inaf.it}

\affiliation[2]{Department of Astronomy, University of Science and Technology of China, Hefei, Anhui 230026, China}
\emailAdd{taoan3068@gmail.com}

\affiliation[3]{Shanghai Astronomical Observatory, Chinese Academy of Sciences, 80 Nandan Road, Shanghai 200030, China}

\affiliation[4]{Leiden Observatory, Leiden University, Einsteinweg 55, 2333 CC Leiden, The Netherlands}
\emailAdd{petley@strw.leidenuniv.nl}

\affiliation[5]{Key Laboratory of Particle Astrophysics, Institute of High Energy Physics, Chinese Academy of Sciences, Beijing 100049, China}
\emailAdd{ailing.wang.wal@gmail.com}

\affiliation[6]{INAF - Istituto di Radioastronomia, via P. Gobetti 101, Bologna, I-40129, Italy}

\affiliation[7]{Physics Department, Technion, Haifa, Israel}
\emailAdd{ behar@physics.technion.ac.il, sina.chen@campus.technion.ac.il, laor@physics.technion.ac.il}

\affiliation[8]{Jodrell Bank Centre for Astrophysics, Alan Turing Building, School of Physics and Astronomy, University of Manchester, Manchester M13 9PL, UK}
\emailAdd{emmanuel.bempong-manful@manchester.ac.uk}

\affiliation[9]{School of Physics, University of Bristol, Tyndall Avenue, Bristol BS8 1TL, UK}

\affiliation[10]{Xinjiang Astronomical Observatory, CAS, 150 Science-1 Street, Urumqi 830011, China}
\emailAdd{cuilang@xao.ac.cn, changning@xao.ac.cn}

\affiliation[11]{Institute of Astronomy, Faculty of Physics, Astronomy and Informatics, NCU, Grudzi\k{a}dzka 5/7, 87-100, Toru\'n, Poland}

\affiliation[12]{Astrophysics Science Division, NASA Goddard Space Flight Center, Greenbelt, MD 20771, USA.}
\affiliation[13]{Center for Space Science and Technology, University of Maryland Baltimore County, 1000 Hilltop Circle, Baltimore, MD 21250, USA.}

\affiliation[14]{Center for Research and Exploration in Space Science and Technology, NASA/GSFC, Greenbelt, Maryland 20771, USA}

\affiliation[15]{Department of Physics, University of Maryland, Baltimore County, 1000 Hilltop Circle, Baltimore, MD 21250, USA}

\affiliation[16]{Department of Physics and Astronomy, The University, Southampton SO17 1BJ, UK}

\affiliation[17]{Instituto de Astrof\'isica de Andaluc\'ia (IAA-CSIC),
Glorieta de la Astronom\'ia s/n, E-18008 Granada, Spain}
\affiliation[18]{School of Sciences, European University Cyprus, Diogenes street, Engomi, 1516 Nicosia, Cyprus}

\affiliation[19]{Department of Astronomy, University of Geneva, ch. d'Ecogia 16, 1290, Versoix, Switzerland}

\affiliation[20]{Kavli Institute for Astronomy and Astrophysics, Peking University, Beijing 100871, China}

\affiliation[21]{Jodrell Bank Centre for Astrophysics, School of Physics and Astronomy, The University of Manchester, Manchester, M13 9PL, UK}

\affiliation[22]{Joint Institute for VLBI ERIC (JIVE), Oude Hoogeveensedijk 4, 7991 PD Dwingeloo, The Netherland}

\affiliation[\dagger]{Chapter co-ordinator}

\abstract{The accretion-ejection mechanism in Active Galactic Nuclei (AGN) remains a central open problem in astrophysics, tied to the role of AGN feedback in galaxy formation and evolution. Radio-quiet AGN dominate the observed AGN population. Lacking luminous jets, their radio emission traces a rich set of processes spanning the host galaxy kpc scales down to the vicinity of the supermassive black hole: star formation, AGN-driven winds and shocks, free-free emission from photo-ionized gas, low-power jets, and coronal activity close to the inner accretion disk. The Square Kilometre Array (SKA) will probe these processes across a wide frequency range with unprecedented sensitivity, wide-field survey capability, and, critically, high-resolution VLBI imaging. Flux, spectral, and polarization monitoring will constrain dynamics and environmental coupling, while mapping nuclear regions on sub-pc to kpc scales will disentangle compact cores from host emission, resolving the diversity of radio activity across accretion regimes and jet powers from the local Universe to the cosmic dawn. At the full AA4 deployment, the SKA-MID phased into global VLBI arrays will deliver sub-milliarcsecond imaging and $\mu$Jy sensitivity over 0.35–15\,GHz, enabling the first population-level census of radio-quiet AGN nuclei. Earlier AA* operations will support pilot studies of the brightest nearby systems.
}

\begin{document}
\maketitle

\section{Introduction}

Active Galactic Nuclei (AGN) are among the most energetic long-lived sources in the Universe, powered by the accretion of matter onto Super Massive Black Holes (SMBHs) at the center of host galaxies. This accretion process releases vast amounts of energy across the entire electromagnetic spectrum, often profoundly shaping the evolution of the host galaxy through AGN feedback \citep[{e.g.},][]{2013ARA&A..51..511K,2014ARA&A..52..589H}.

AGN have been historically classified based on the relative strength of their radio emission. The classic radio dichotomy was first formalized by \citet{1989AJ.....98.1195K}, who found a clear bimodal distribution in the ratio of radio to optical luminosity ($R \equiv f_{5\,\mathrm{GHz}}/f_{B}$). This divides the population into Radio-Loud (RL, $\sim 10\%$, $R\gtrsim10$) and Radio-Quiet AGN (RQ, $\sim 90\%$, $R\lesssim 10$). RL AGN launch powerful, long-lived, highly collimated relativistic jets that frequently 
inflate kpc–Mpc radio lobes, show strong Doppler boosting, and can carry kinetic powers of 
$\sim10^{44}$–$10^{46}\ \mathrm{erg\ s^{-1}}$ \citep{1989AJ.....98.1195K,2014ARA&A..52..589H,2013ARA&A..51..511K,2007ApJ...658..815S,1977MNRAS.179..433B}. In contrast, RQ AGN typically host compact (sub-pc to kpc), lower-power outflows with mildly or intermittently relativistic motion (see Figure 1) \citep{2025ApJ...987L..26W, 2023MNRAS.523L..30W}, weaker Doppler boosting \citep{2007ApJ...658..815S}, lower brightness temperatures ($T_{\mathrm{B}}\!\lesssim\!10^{7\text{–}9}$ K) \citep{Nagar2002,Wang2023paperI,Wang2023paperII,chen25}, and little or no large-scale lobe development \citep{1984ApJ...285..439U,1989AJ.....98.1195K,gallimore2004}.

Although radio loudness is a convenient empirical metric, the spread across the AGN population more likely reflects variations in jet production efficiency, magnetic-flux supply, and
accretion state, with Doppler boosting and host contamination further sculpting the observed distribution into an apparent bimodality \citep{2007ApJ...658..815S,Tchekhovskoy2011,2012MNRAS.421.1569B,Panessa2019, An2026}. A complementary jetted/non-jetted classification has been advocated in which sources are separated by the presence of a
resolved, high-brightness-temperature jet rather than by a luminosity ratio \citep{Padovani2017}.  
A time-domain radio-loudness (TDRL) framework has recently been proposed to recast this dichotomy as an epoch-resolved, activity-state-dependent quantity, capturing sources whose apparent loudness migrates across the conventional boundary over an AGN duty cycle \citep[][in review]{An2026}. Throughout this chapter we retain the traditional $R<10$ designation, noting that part of the RQ population is genuinely non-jetted, even at milliarcsecond angular resolution, while another part harbours jets that are confined, intermittent, or obscured.

Understanding the origin of the faint radio emission from RQ AGN is essential for a coherent picture of galaxy evolution. The RQ population makes up the vast majority of AGN, implying that their low-level radio emission is tied to the most common mode of SMBH accretion and feedback \citep[e.g.][]{Kharb23}. The core challenge, therefore, lies in spatially, spectrally, and temporally resolving the weak signal that blends weak jets, disk–corona processes, winds, and circum-nuclear star formation into a single nuclear component \citep{Panessa2019}.

The most straightforward hypothesis is that RQ AGN still launch scaled-down relativistic jets. 
Lower jet power can result from limited magnetic-flux accumulation onto the black hole
and from accretion states that increase mass loading or reduce magnetization, even if the spin plays a non-negligible role \citep{1977MNRAS.179..433B,1995ApJ...452..710N,Tchekhovskoy2011,2012MNRAS.423.3083M,2007ApJ...658..815S,2007ApJ...667..704V}. 
Intermittent fueling and short duty cycles can also yield compact, transient ejections without sustained lobe inflation \citep{Hardcastle01.2026.SKA}. In ``frustrated'' or ``choked'' scenarios, a dense, gas-rich circum-nuclear medium  decelerates and disrupts the flow, with additional free–free absorption suppressing extended emission, so that the radio output  remains confined to sub-parsec to parsec scales \citep{1996ApJ...458..136G,1998PASP..110..493O, Mukherjee2018, 2021MNRAS.504.3823W}.
Such compact, persistent structures carry the signature of a self-absorbed synchrotron core with flat or inverted spectral index ($\alpha \gtrsim -0.5$, F$_{\nu}$ $\sim$ $\nu^{\alpha}$)  (e.g., \citealt{baldi22,chen25}).

A second channel invokes powerful, non-relativistic outflows or winds launched from the accretion disk or the inner broad-line region. These high-velocity flows interact with the surrounding interstellar medium (ISM), generating shocks that  accelerate relativistic electrons (e.g., \citealt{Faucher2012}) and producing diffuse, steep-spectrum synchrotron radiation ($\alpha \approx -0.7$) potentially resolved on hundred-parsec scales, thereby serving as a direct tracer of large-scale AGN kinetic feedback (e.g., \citealt{Harrison2018})

Another possible explanation for the observed radio signal is related to the intense star formation (SF) activity within the host galaxy (e.g., \citealt{Padovani2016}). Radio emission from supernova remnants and H\textsc{ii} regions is a well-established tracer of the Star Formation Rate (SFR), typically following a strong Far-Infrared (FIR)--radio correlation \citep[e.g.,][]{Condon1992, Sargent2010}. High angular resolution is necessary to disentangle this diffuse, galactic-scale emission from a point-like nuclear source \citep{kharb06,baldi21, Cheng2025}.

A physically distinct mechanism attributes the radio signal to the hot, magnetically dominated X-ray emitting corona located in the innermost regions close the black hole. This scenario suggests that magnetic reconnection events within the AGN corona can accelerate relativistic electrons, producing compact, highly variable, low-level radio emission \citep[e.g., ][]{LaorBehar2008}. The emission may be 
%free-free (thermal) or 
gyro-synchrotron (non-thermal) radiation 
\citep[e.g., ][]{Panessa2019}, typically characterized by short-timescale flaring \citep{Shablovinskaya2024}, which distinguishes it from the steady-state emission of mini-jets or large-scale winds. 
Because it is anchored to the immediate SMBH environment, the coronal component offers a rare probe of the magnetic fields and energy dissipation close to the black hole event horizon. The millimeter-wave core emission, typically in the $100-300\ \mathrm{GHz}$ range, is tightly correlated with X-ray luminosity, supporting this scenario \citep{Baldi2015, petrucci23, ricci2023}.

The prevalence of the RQ population is not constant across cosmic time. Observational evidence suggests that the comoving density of all AGN peaks at around redshift $z \sim 2$ (the cosmic noon), tracing the peak of star formation activity and black hole accretion  \citep{2009MNRAS.399.1755C,2010MNRAS.401.2531A,2014ARA&A..52..415M}. However, the precise evolutionary path and the physical processes driving the RQ fraction relative to the RL AGN, and their respective contributions of each to heating the circum-galactic medium, remain poorly constrained, particularly at high redshifts \citep{2007ApJ...656..680J,2015AJ....149...61K,2024MNRAS.528.5692K}. 
Since RQ AGN dominate the accreting black-hole budget over most of cosmic history, quantifying the origin of their radio emission is indispensable for assessing their true impact on galaxy evolution and the quenching of star formation \citep{2008MNRAS.388.1011M,2017A&ARv..25....2P, Alexander2025, Prandoni01.2026.SKA}.

The SKA, and especially its Very Large Baseline Interferometry (VLBI) mode, will deliver the sensitivity and resolution needed to break the present degeneracies between jet, wind, star-formation, and coronal origins of RQ radio emission. 
In this chapter, we outline the scientific capability of the SKA in this field, spanning a wide range of wavelengths and featuring strong performance, particularly as a VLBI element, the SKA will allow for a step change in our understanding of the origin and cosmic evolution of accretion and ejection in RQ AGN. The remainder of the chapter is organised as follows. Section~\ref{sec:local_universe} develops the accretion--ejection physics accessible to SKA--VLBI in the local Universe, with emphasis on brightness temperature, spectral, core shift and polarimetric discriminants that separate jet, corona, wind and star-formation origins. Section~\ref{sec:high_z} extends the discussion to the early Universe, covering high-redshift quasars, the RQ AGN population identified in deep radio surveys, and the super-Eddington regime. Section~\ref{sec:conclusions} summarises the expected outcomes and the synergies with facilities operating in other wavebands.

 At the AA4 stage, SKA-MID will operate with four VLBI-capable bands (1, 2, 5a, 5b) and provide tied-array beams feeding intercontinental baselines with the EVN and African partners; the Australian Long Baseline Array (LBA) provides the complementary southern-hemisphere coverage that is in fact geometrically better matched to the SKA--MID site, and we anticipate SKA--MID tied-array participation in LBA sessions to be a core mode of operation for RQ AGN at southern declinations. This configuration provides sub-milliarcsecond (sub-mas) resolution and $\mu$Jy imaging sensitivity.
 The earlier AA* phase will allow focused observations of the brightest Seyferts and quasars, serving as a technical and scientific pathfinder for the full AA4 programme.

\section{Accretion--ejection physics in the local Universe with SKA--VLBI capabilities}
\label{sec:local_universe}

Prior to the advent of the SKA, interferometric observations of RQ AGN utilizing current facilities (e.g., VLA, e-MERLIN, GMRT, ATCA, EVN, VLBA, and LOFAR) have provided key constraints, yet have consistently highlighted the fundamental limitations in sensitivity and resolution needed to definitively disentangle emission sources \citep[e.g.,][]{Panessa2019, Sebastian2020, Silpa2021a, Silpa2021b, Silpa2022, baldi22, Kharb23, chen25}.

VLBI studies of local Seyfert galaxies have established the ubiquity of compact sub-parsec cores with $T_{\rm B}\gtrsim 10^{6-7}$\,K \citep{Ulvestad1995, Ulvestad2001,Wang2023paperI, Wang2023paperII,Cheng2025}. In many cases, hints of parsec-scale jets or misaligned structures point to precession or jet--ISM interaction \citep{Giroletti2009,Panessa2013,2021MNRAS.504.3823W}. In complete nearby samples, most VLA-detected nuclei are also detected with VLBI at 1.7 and/or 5\,GHz, with typical core powers $\log P_{5\,{\rm GHz}}\sim 19.4$\,W\,Hz$^{-1}$ and characteristic sizes of order $0.05$\,pc at 10\,Mpc \citep{Nagar2002, Nagar2005, Panessa2013, Nyland2016,Cheng2025}. 
These studies reveal a diversity of mechanisms, from jet-base synchrotron to steeper, shock-accelerated components. SKA--VLBI will systematically quantify their relative contributions across volume-limited samples rather than the handful of bright cases accessible today.

The nearby Universe provides the cleanest laboratory for disentangling the weak nuclear radio emission of RQ AGN from circum-nuclear star formation and galactic-scale outflows. The critical missing capability has been the simultaneous provision of milli-arcsecond (mas) angular resolution to isolate the sub-parsec nucleus, $\mu$Jy\,beam$^{-1}$ sensitivity to detect compact components with brightness temperatures above the thermal regime, broad frequency coverage to separate synchrotron self-absorption from free--free and to perform Faraday diagnostics, and time-domain stability to track state changes in the accretion flow and flaring activity in the hot corona.

SKA--MID in its AA4 configuration, phased for integration into global VLBI networks, delivers exactly this combination: four initially deployed receiver bands (Bands~1, 2, 5a, 5b) spanning $0.35$--$15.4$\,GHz, connected-element baselines up to $\sim 150$\,km, and critically, tied-array beams that transform SKA-MID into an ultra-sensitive VLBI element on intercontinental baselines, delivering $\sim 1$\,mas at $8$\,GHz and better than $0.5$\,mas in Band 5b (8.3 -- 15.4\,GHz), providing linear scales of $\sim 0.1$--$1$\,pc at the typical distances of nearby galaxies ($z\lesssim 0.1$).

At these resolutions and sensitivities, the brightness temperature ($T_{\rm B}$) becomes the primary physical discriminant. Adopting a circular Gaussian source profile, a 30\,$\mu$Jy core at 10\,GHz observed with a 0.5\,mas beam yields $T_{\rm B}\approx 1.5\times 10^{6}$\,K, safely above H\,\textsc{ii} / free--free values and in the regime of compact synchrotron emission; even a $20\,\mu$Jy component at $5$\,GHz with a $1$\,mas beam yields $T_{\rm B}\approx 10^{6}$\,K. With SKA--VLBI routinely reaching ${\rm rms}\lesssim 1\,\mu$Jy\,beam$^{-1}$ in $8$\,hrs, $5\sigma$ detections of $5$--$10\,\mu$Jy cores are feasible, enabling the transition from mJy-biased pilot studies to a genuine census of nuclear radio components in intrinsically RQ systems.

\subsection{\textbf{SKA--VLBI capabilities to disentangle different radio emission mechanisms}}

Beyond detection, broadband imaging with the SKA Band~5 (4.6 -- 15.4\,GHz) coupled to Band~2 ($\simeq$1.3 -- 1.7\,GHz) provides spectral, opacity, and Faraday diagnostics central to classification (see Table \ref{tab:skao_diagnostics_condensed}). Exploiting the full scientific potential of this broadband capability demands that the partner VLBI stations (EVN, VLBA, African VLBI Network, LBA) deliver contemporaneous coverage matched to the SKA--MID frequency range. This carries concrete programmatic implications for the upgrade path of the non-SKA elements, and we flag it here as a prerequisite for core-shift and Faraday tomography at the precision discussed below. Compact jet bases are expected to show flat or slightly inverted spectra at cm wavelengths due to synchrotron self-absorption, with frequency-dependent core positions $ r_{\rm core}(\nu)\ \propto\ \nu^{-1/k}$, encoding particle density and magnetic-field gradients near the launch region \citep{Lobanov1998}. Measurement of core shifts between $8$ and $15$\,GHz at the level of tens of $\mu$as is realistic with phase-referenced SKA--VLBI, allowing direct estimates of the magnetic field ($B$) and pressure near $\sim 10^{3-4}\,R_{\rm g}$ (where $R_{\rm g}=GM_{\bullet}/c^{2}$ is the gravitational radius). The same wide-band datasets, recorded in full Stokes, yield rotation-measure (RM) maps and polarization morphologies that distinguish ordered fields in a collimating jet from the patchier, more depolarizing patterns associated with wind shocks \citep{Hovatta2012,Mahmud2013}.

Time-domain capability will be transformative for identifying transient jet or coronal events. The well-established radio--X-ray correlation in RQ quasars, $L_{\rm R} \sim 10^{-5}\,L_{\rm X}$ (the AGN analogue of the G\"udel--Benz relation), argues that at least part of the RQ nuclear emission arises in a magnetized disk corona \citep{LaorBehar2008,Behar2015,Behar2018,Panessa2019}. Coordinated, high-cadence SKA--VLBI monitoring during X-ray bright phases can test whether radio-quiet nuclei display the same disc--corona--jet coupling phenomenology seen in radio-loud objects and X-ray binaries: core brightening and spectral hardening during the rising phase, followed by the appearance of new mas-scale components and evolving polarization as energy is redistributed. Correlating SKA--VLBI variability and core-shift evolution with X-ray spectral state transitions will discriminate between corona-dominated emission and contributions from low-power jets or winds.
In addition, high-cadence monitoring campaigns with sufficient sensitivity to detect rapid, low-level flaring from magnetic reconnection events in the accretion disk corona will open a largely unexplored time-domain window on RQ nuclear activity.

A critical confounding factor at arcsecond resolution is host-galaxy emission: diffuse synchrotron from star formation follows the FIR--radio correlation and can dominate the total 1--3\,GHz flux \citep{Condon1992}. SKA--VLBI overcomes this by resolving out the kpc-scale disk and recovering the nuclear SED at mas scales; matched-$uv$ connected-element imaging at sub-arcsecond resolution provides stringent host subtraction. Where the residual, resolved-out emission follows the FIR--radio relation, we attribute it to star formation; where it correlates with optical outflows, steep spectra, and edge-brightened kinematics, we identify AGN-driven wind shocks as the dominant extended component \citep{Harrison2018, Wang2023paperII}. In either case, nuclear compactness and $T_{\rm B}$ remain the definitive indicators  of an AGN origin.

The wind-shock interpretation carries broader physical significance because energy-conserving outflows efficiently convert a fraction of the AGN luminosity into mechanical work, boosting momentum flux on kpc scales and producing steep-spectrum radio emission at outflow edges \citep{Harrison2014,Morganti2015,Jarvis2019}. Spatially resolved studies with VLA, MeerKAT, and e-MERLIN have revealed diffuse, polarized structures co-spatial with ionized or molecular outflow cones in several nearby Seyferts and quasars, showing rotation-measure (RM) and depolarization behavior distinct from collimated jet bases \citep{Gallimore2006,BarcosMunoz2018,Mingo2022}. The Quasar Feedback Survey has proven especially powerful in this regard, combining spatially resolved radio continuum, polarimetry, and optical integral-field data to isolate jet-, wind-, and star-formation-dominated components across a homogeneous sample of obscured $z\lesssim 0.2$ AGN \citep{Jarvis2019,Silpa2022}; the survey's rotation-measure results demonstrate that depolarization asymmetries and RM gradients can unambiguously separate collimated low-power jets from mass-loaded wind shocks in sources even when the total-intensity morphology alone remains degenerate \citep{Sebastian2020,Silpa2021a,Silpa2021b,Kharb23}. If such signatures proven common among local RQ AGN, winds must account for a significant fraction of their radio power budget.

\begin{figure}[h!]
    \centering
    \includegraphics[width=0.9\linewidth]{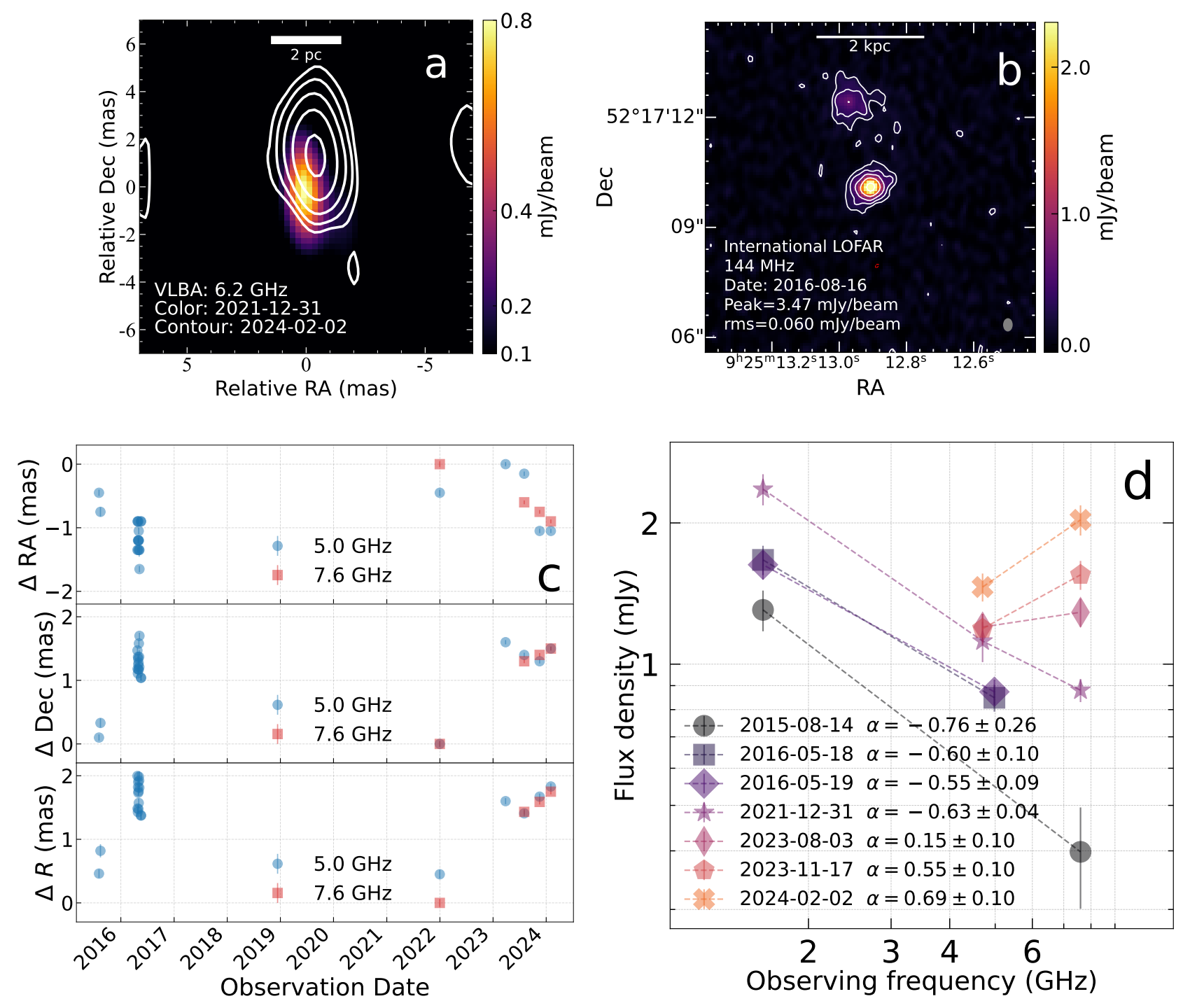}
    \caption{VLBI evidence that a subset of RQ AGN launches relativistic jets, illustrated by Mrk~110 as a representative case (taken from \citet{2025ApJ...987L..26W}).
    (a) VLBA 6.2\,GHz images on 2021 December 31 (colour scale) and 2024 February 2 (contours), revealing a core offset of $\sim 1.6$\,mas. The image centre is set at the 2021 December peak position (R.A. 09:25:12.84781, Dec. $+52$:17:10.3862). (b) International LOFAR image at 144\,MHz observed on 2016 August 16, revealing a $\sim 1.7$\,kpc northern extension aligned with the parsec-scale jet. (c) Peak-position evolution. (d) Radio-spectrum evolution from 2015 to 2024, documenting a transformation from a steep ($\alpha=-0.76$) to an inverted ($\alpha=+0.69$) spectral index. Historical data from 2015--2016 are from \citet{2022MNRAS.510..718P}; dashed lines are power-law fits. This source exemplifies the class of measurement (compactness, core shift, spectral inversion, and proper motion) that SKA--VLBI will deliver for volume-limited samples, extending such diagnostics well beyond the few bright cases currently accessible.}
    \label{fig:Mrk110}
\end{figure}

%% Table rotated to landscape for legibility (Referee 1 request).
\begin{sidewaystable}
\centering
\small
\caption{\textbf{SKA--VLBI diagnostics for radio-quiet AGN emission mechanisms.}}
\label{tab:skao_diagnostics_condensed}
\begin{tabular}{|c|p{5cm}|p{5cm}|p{6cm}|}
\hline
\textbf{Mechanism} & \textbf{Primary Physical Origin} & \textbf{Characteristic Observational Signatures (Pre-SKA)} & \textbf{Key SKA--VLBI Requirements \& Estimates} \\
\hline
\textbf{Jets/Jet base} & Scaled-down, mildly relativistic, collimated outflow from the accretion disk base. & \textbf{Morphology}: Compact core, unresolved, or core-jet structure ($<1 \text{ pc}$). \textbf{Spectrum}: Core: Flat or inverted ($\alpha \gtrsim -0.5$), SSA emission. Jet: optically thin steep spectrum. \textbf{Kinematics}: Proper motion detected & \textbf{Resolution}: $\lesssim 1 \text{ mas}$ imaging. \textbf{Sensitivity}: $\sim$ $\mu \text{Jy}$ (AA4) to detect faint cores. \textbf{Diagnostic}: Measure knot motions, map polarization structure \\
\hline
\textbf{Corona} & Non-thermal/thermal processes (magnetic reconnection) in the hot, compact accretion disk corona. & \textbf{Morphology}: Ultra-compact ($<0.1 \text{ pc}$), unresolved core. \textbf{Spectrum}: Flat/inverted ($L_R/L_X \sim 10^{-5}$). \textbf{Kinematics}: Expected rapid, non-steady flaring/variability. & \textbf{Resolution}: Required $\lesssim 1 \text{ mas}$ resolution to isolate the compact source. \textbf{Sensitivity}: High cadence monitoring; $\mu \text{Jy}$ detection for variability. \textbf{Diagnostic}: Simultaneous X-ray/radio monitoring to test the Neupert effect.
\\
\hline
\textbf{Winds} & Synchrotron emission from shocks generated as an uncollimated AGN outflow interacts with the ISM. & \textbf{Morphology}: Diffuse, irregular structures, extending $\sim 100 \text{ pc}$. \textbf{Spectrum}: Steep ($\alpha \approx -0.7$), optically thin. \textbf{Kinematics}: Slow bulk speeds. & \textbf{Resolution}: High mas-resolution needed to resolve the outflow  base  from the jet region.  \textbf{Sensitivity}: $\sim 2\,\mu$Jy\,beam$^{-1}$ (SKA--Low) for faint relic/shocked plasma. \textbf{Diagnostic}: Polarization mapping (Faraday RM). \\
\hline
\textbf{Star Formation} & Diffuse synchrotron emission from supernova remnants and thermal free-free from HII regions. & \textbf{Morphology}: Diffuse, host-like (kpc scales). \textbf{Spectrum}: Steep ($\alpha \approx -0.7$), matching the FIR-radio correlation. \textbf{Kinematics}: Non-variable; highly depolarized. & \textbf{Resolution}: $\lesssim 1 \text{ mas}$ resolution is necessary  to resolve out  the extended background. \textbf{Sensitivity}: $\sim$ $\mu \text{Jy}$ sensitivity ensures the faint nuclear component is cleanly isolated. \textbf{Diagnostic}: Spectral index mapping and spatial correlation with FIR tracers. \\
\hline
\end{tabular}
\end{sidewaystable}

At mas scales, SKA--VLBI resolution can resolve the base of wide-angle outflows on parsec scales, providing direct observational benchmarks for numerical simulations of line-driven or magnetic winds. This high resolution will help distinguish between a persistent jet and the initial expansion of a nuclear wind/shock region. Moreover, the SKA offers excellent polarimetric capabilities that can reveal the ordered magnetic fields near the SMBH, a governing factor in jet formation. Mas-scale polarization mapping of the compact cores will constrain   the magnetic field topology responsible for launching the jet or wind.

The SKA's superior polarimetric capabilities  (rms noise less than a few $\mu$Jy~beam$^{-1}$ in full Stokes) will allow enable detailed mapping of the strength and geometry of magnetic fields in the circumnuclear medium. Synchrotron emission generated by wind-shock acceleration is expected to be co-spatial with known outflows (traced by multi-wavelength data) but exhibit a distinct polarization morphology and magnetic field structure compared to highly organized, relativistic jets \citep{Alexander2025, Kudoh01.2026.SKA}.

The contrast in radio polarization offers a key discriminant between physical mechanisms powering the sub-parsec core. Flat-spectrum, gr-$T_{\rm B}$ cores exhibiting that exhibit negligible or extremely low polarization levels constitute the expected accretion-disk corona model. This low polarization results from the emission being generated in an isotropic or quasi-spherical region, leading to the cancellation of the total electric vector polarization across different viewing angles \citep{LaorBehar2008,Behar2015,InoueDoi2018,baldi21,Chen2023, Wang2023paperI, Wang2023paperII}.

Conversely, the observation of cores displaying detectable, ordered polarization (even if low) and systematic core shifts would favor a low power jet model. These features are characteristic of synchrotron self-absorbed jet bases, where the magnetic field structure is ordered, and the emission is produced in a collimated outflow \citep{Panessa2019}. \citet{Sebastian2020} and \citet{Silpa2021b} provide observational support for this discriminator by recovering ordered polarization structures and non-zero rotation measures in compact, flat-spectrum Seyfert cores where the coronal model would predict negligible polarization.

A single SKA--VLBI observation simultaneously delivers mas-scale morphology, a spectral index spanning Band~2 to Band~5a, and full-Stokes polarimetry; only when all three diagnostics are available in concert can the coronal and low-power-jet hypotheses for the compact nucleus of a radio-quiet AGN be reliably separated. The power of this approach rests not on the production of a single dataset \textit{per se}, but on the fact that the three discriminants carry independent systematics and respond differently to source state, so that cross-comparison eliminates degeneracies that any individual diagnostic would leave unresolved.

Operationally, the SKA--VLBI strength lies in a stable, wideband, full-polarization calibration and the ability to phase SKA--MID into EVN and VLBA sessions, with the LBA providing the analogous function for southern targets. Modern, reproducible pipelines such as \textsc{rPICARD} and \textsc{GPCAL} have been validated on comparable broadband datasets and support absolute EVPA tying across bands \citep{Janssen2019,Kim2024}. With phase referencing to nearby calibrators and contemporaneous multi-frequency registration, positional shifts of order $10$--$50\,\mu$as can be recovered across 5--8\,GHz where the partner VLBI arrays provide adequate baseline coverage, permitting magnetic-field and pressure gradients to be inferred near the jet base.

A natural observing program is a volume-limited ($D\lesssim 100$\,Mpc) sample of nearby Seyferts and high-$L/L_{\rm Edd}$ RQ AGN observed with SKA--VLBI at Band~5 for the core/jet base and polarization, and Band~2 for opacity and RM, with $8$--$10$ epochs over a year and $\mu$Jy-level sensitivity per epoch. For typical $20$--$50\,\mu$Jy cores at 5--10\,GHz, we expect $T_{\rm B}\sim 10^{6}$\,K at 1\,mas and $\sim4 $ times higher at 0.5\,mas, ensuring separation from thermal processes. Simultaneous X-ray monitoring tests whether radio/opacity changes track coronal state, while ALMA/mm monitoring probes the putative coronal synchrotron component that often peaks near $100$\,GHz \citep{Baldi2015,Behar2015, Behar2018, InoueDoi2018,petrucci23, Shablovinskaya2024, Mutie2025, delPalacio2025}. Detections of core shifts at the $30$--$100\,\mu$as level between 8 and 15\,GHz, if accompanied by spectral flattening and ordered polarization during bright phases, would constitute evidence for magnetically dominated launch episodes. Strong depolarization with enhanced RM and stalled morphology would favor mass-loaded winds. Either outcome tightens the constraints on disc--jet (or disc--wind) coupling in the RQ regime.

Finally, SKA--VLBI enables a clean test of whether the RQ population adheres to the black-hole fundamental plane that relates radio luminosity, X-rays, and mass \citep{Merloni2003, Saikia2018, Ruffa2023}.This relationship appears to hold down to stellar-mass black holes, but  the scatter can be significantly reduced by assembling  a large, well-defined and unbiased sample of radio detections of low-mass SMBHs  associated with RQ AGN \citep{Plotkin2012, Gultekin2022, Wang2024}. By delivering nuclear-only radio measurements with known $T_{\rm B}$ and morphology, SKA--VLBI can either confirm that RQ nuclei obey the same accretion--ejection scalings as their radio-loud counterparts, or reveal  a genuine bifurcation tied to magnetization and launch efficiency at low jet powers.

\section{Radio activity in the early Universe}\label{sec:high_z}

Understanding RQ AGN beyond $z\sim3$ is key to tracing the evolution of low-power jets and magnetic flux across cosmic time.
Since the angular diameter distance reaches a maximum around $z\approx1.5$, the power of VLBI to study AGN accretion and ejection physics on sub-galaxy scales at high redshift is only limited by the luminosity of the sources available to us and the sensitivity of our telescopes. The AA4 array will allow observations with SKA--MID at a resolution ranging from 0.3$^{\prime\prime}$ to 0.03$^{\prime\prime}$, while for SKA-LOW the range is 10$^{\prime\prime}$ to 2$^{\prime\prime}$. When incorporated into a VLBI network, SKA-MID achieves image-plane resolutions of 1--10\,$\mu$as as at rms sensitivities of a few $\mu$Jy\,beam$^{-1}$\citep{li_vlbi_2024}. The challenge for VLBI SKA-LOW is more significant, since fewer sites operate globally at these frequencies; however, with the potential to study intrinsically fainter sources, thanks to the synchrotron power law slope, it could still be crucial to the study of RQ AGN at high redshift \citep{Kondapally01.2026.SKA, Spingola01.2026.SKA}. SKA--LOW offers enormous survey speed but faces a confusion-limited floor at arcminute-to-arcsecond scales unless very long baselines are available; consequently, SKA--LOW will be superb for demographics and spectral curvature, while SKA--MID + VLBI provides the compactness, brightness temperature, polarization, and core-shift diagnostics needed to isolate the nucleus.

Separating radio emission from star formation and AGN activity will become an ever greater challenge with the SKA. Particularly for SKA-LOW, without an improvement in resolution, AGN studies will be severely limited by confusion noise, and it will be difficult to progress past LOFAR deep surveys apart from surveying more quickly and in the southern hemisphere.

\subsection{Quasars}

For luminous quasars at $z \gtrsim 6$, the most open issues are the prevalence and duty cycle of relativistic outflows at the lowest detectable powers, and the physical origin of the faint radio population that dominates deep counts. Current constraints on the radio-loud fraction at the highest redshifts remain limited by sensitivity and sample size \citep{Banados2015, Perger2017}. Existing VLA surveys produce few detections at $\sim$100–200\,$\mu$Jy and large uncertainties, while VLBI detections of individual sources confirm compact, high-$T_{\rm B}$ cores consistent with jet bases \citep{An2020,An2023,Belladitta2023,Belladitta2020,Frey2011,Frey2024,LiuYQ2022a,LiuYQ2022b,Liu2024,Ighina2021,Momjian2018,Sbarrato2012,Spingola2020,radcliffe21}. 

SKA refines the question from ``\emph{are jets present?}'' to ``\emph{how do they couple to the flow and environment?}''. Not going into details about sensitivity limits here, we emphasize the diagnostics that uniquely matter at high $z$. Nuclear compactness and spectral shape on parsec scales reveal the nature of the central engine, while wide-band polarization and Faraday rotation measures trace magnetized screens threading the inner few hundred parsecs. Multi-frequency phase registration further recovers \emph{core shifts} that encode opacity and $B$-field gradients along the jet, directly testing launching scenarios under the intense radiative and IC/CMB losses expected at these redshifts \citep{Ghisellini2014, Spingola01.2026.SKA, Kondapally01.2026.SKA}. Paired LOW+MID connected-element imaging then ties the parsec-scale core to any steep, edge-brightened structures associated with quasar-driven winds, placing the faint radio output of the majority population on a firm physical footing. 

Beyond direct source characterization, compact high-$z$ quasars selected and characterized by SKA--VLBI provide the bright, stable backlights needed for HI~21-cm absorption experiments and tomography during reionization, while their RM and depolarization patterns probe the early magnetized CGM/IGM \citep{Miley2008,Mesinger2016}. In combination with JWST/ALMA constraints on host gas and star formation, SKA will establish whether the typical high-$z$ quasar is best described by a weak, intermittently mass-loaded jet (or wind) anchored in a magnetized corona, or by truly jetless accretion whose radio output is dominated by wind shocks and circumnuclear star formation. This distinction is only accessible with the multi-scale, polarization-aware view that the SKA provides.

\subsection{Radio-Quiet AGN}
The prospect of studying RQ AGN is bolstered by the classification of sources in deep radio images taken by SKA pathfinders. For example, \cite{best_lofar_2023} classified $\sim 80,000$ sources in the first LoTSS: Deep Fields data release using several SED fitting codes to find star-forming galaxies, Low Exitation Radio Galaxies, High Exitation Radio Galaxies and RQ AGN. They found that the RQ AGN fraction grows from less than 10\% at $z=1$ to more than 30\% by $z=4$ while other populations decrease. Given that the multi-wavelength information needed to estimate star formation rates will become sparser at high-redshift, brightness temperature measurements using SKA--VLBI may become the only way to separate AGN and star formation radio emission for these distant radio-quiet sources.

Extrapolations from LoTSS-Deep and MIGHTEE luminosity functions indicate that, within a 100\,deg$^{2}$ SKA--LOW field reaching 5\,$\mu$Jy\,beam$^{-1}$, several $\times 10^{4}$ radio-quiet AGN will be detected up to $z\sim 4$ \citep{Hale2025}, with a few hundred compact enough for VLBI follow-up.
Such statistics will, for the first time, enable population-level tests of jet duty cycles, magnetic-field evolution, and feedback efficiency across cosmic time \citep{Hardcastle01.2026.SKA}.

\section{Conclusions and Future Synergies for RQ AGN Research}\label{sec:conclusions}

The field of RQ AGN is poised for a major transformation, moving beyond the current limitations imposed by low sensitivity and insufficient angular resolution. The fundamental ambiguity, whether the core radio emission originates from scaled-down relativistic jets, a hot magnetized accretion disk corona, or kinetic wind shocks, will be resolved by the SKA. SKA--VLBI is the key differentiator here, as it enables a statistically complete census of nuclear components. The SKA-Mid AA4 configuration will achieve $\sim 1$\,mas imaging at $5$\,GHz with a thermal rms of a few $\mu$Jy\,beam$^{-1}$\,hr$^{-1}$ (deeper limits are reached only in long integrations), providing the resolution to measure knot motions and derive fundamental properties like the magnetic field and degree of polarization on parsec scales. This kinematic and spectral precision is essential for distinguishing true jet activity from coronal emission. Furthermore, the exquisite sensitivity of SKA-VLBI polarimetry will allow detailed studies of the magnetic-field structure, directly testing the coronal versus jet launch scenarios. The full power of the SKA is realized through multi-frequency synergies. Coordinated X-ray (e.g., NewAthena, Einstein Probe) and mm/sub-mm (ALMA, ngVLA) monitoring, combined with optical/IR IFU spectroscopy (JWST, MUSE) and optical/NIR light curves (LSST) will tie SKA-VLBI nuclear diagnostics to coronal heating, wind energetics, and host feedback. GRMHD and polarized radiative-transfer simulations will interpret SKA polarization and core-shift data in terms of magnetic flux and mass-loading at the jet base. By applying these multi-domain diagnostics, the SKA will enable the physical classification of the RQ AGN population, advancing the field from conjecture to a quantitative, multi-mechanism understanding of accretion and feedback physics in the dominant class of AGN.

Several plausible enhancements would extend the RQ AGN science case beyond what AA4 already delivers. A Band 6 receiver (or SKA--MID participation in mm VLBI through dedicated front-ends) would connect the self-absorbed cm cores measured here to the optically thin mm regime where the putative coronal synchrotron component peaks, bridging the frequency gap to ALMA at the scale of the jet launch region. Higher-time-resolution voltage-buffer recording would enable sub-second coronal flare studies coordinated with Einstein Probe and NewAthena triggers, probing variability at cadences shorter than the $\sim$few-minute integrations that set the thermal noise floor in standard VLBI. Extended baselines, in particular participation in mm space-VLBI missions or long-baseline additions in Africa, would push the angular resolution into the tens of $\mu$as regime required to resolve the jet base of a nearby Seyfert at $10^{2}$--$10^{3}\,R_{\rm g}$ \citep{Bempong-Manful01.2026.SKA}. Each of these capabilities is an incremental upgrade to the AA4 baseline rather than a replacement, and each targets a specific measurement whose cost--benefit can be evaluated against the first-generation AA4 results.

\subsection*{Acknowledgements}
FP and GB acknowledge financial support from the Bando Ricerca Fondamentale INAF and "Programma
di Ricerca Fondamentale INAF 2023 and 2024.
E.B. is supported by The Israel Science Foundation (grant No. 2617/25).

\bibliographystyle{abbrvnat-maxbibnames4}
\bibliography{chapter}

\end{document}